\documentclass[aps,prl,twocolumn]{revtex4-1}
\usepackage{amssymb}
\usepackage{epsfig}
\usepackage{graphicx}
\usepackage{amsmath}
\usepackage{amsfonts}

\newcommand{\ket}[1]{\left\vert #1 \right\rangle}

\newcommand{\bla}[1]{\left( #1 \right)}
\newcommand{\blb}[1]{\left[ #1 \right]}

\begin{document}

\title{Quantum limit for avian magnetoreception: How sensitive can a chemical compass be?}

\author{Jianming Cai, Filippo Caruso, Martin B. Plenio}

\address{Institute for Theoretical Physics, Albert-Einstein-Allee 11, University Ulm, D-89069 Ulm, Germany}

\begin{abstract}
The chemical compass model, based on radical pair reactions, is a fascinating idea to explain avian magnetoreception. At present, questions concerning the key ingredients responsible for the high sensitivity of a chemical compass and the possible role of quantum coherence and decoherence remain unsolved.  Here, we investigate the optimized hyperfine coupling for a chemical compass in order to achieve the best magnetic field sensitivity. We show that its magnetic sensitivity limit can be further extended by simple quantum control and may benefit from additional decoherence. With this, we clearly demonstrate how quantum coherence can be exploited in the functioning of a chemical compass. The present results also provide new routes towards the design of a biomimetic weak magnetic field sensor.

\end{abstract}

\date{\today}

\maketitle

{\it Introduction.---} Quantum effects in biology has recently raised considerable interest in the communities of physics, biology and chemistry \cite{LHCE,SME1,Ritz00,Sch78,Sol07,Sol10,Ritz10}. Current research is concerned with understanding how long-lived quantum coherence may exist and what role it may play in several important biological processes \cite{LHCT1,Caruso09,Caruso10,Sar10,SME2,Cai10prl,Ved09,Kom09JonHor10}. The development of quantum biology may in turn help us to advance biomimetic quantum technologies. One particular example of quantum biology, in which quantum effects may be relevant, is avian magnetoreception. As one of the main hypotheses to explain the magnetic sensing of many species \cite{Rod09,JohRev}, the radical pair mechanism involves unpaired electrons that have intermediate correlated spin states \cite{Sch78,Ritz00,Sol07,Sol10,Ritz10}. Several important experiments have provided compelling evidence to support the hypothesis that the radical pair mechanism is exploited in avian magnetoreception \cite{Ritz04,Wil96} (although other mechanisms are believed to make contributions as well), and important observations towards the understanding of the physiological basis for avian magnetoreception \cite{Zap09,Mou09}.  (a)	It was argued that quantum entanglement (i.e. non-classical correlations between two radicals) \cite{Chuang} and interaction with environment may be important for  the functioning of a chemical compass \cite{Cai10prl,Ved09,Kom09JonHor10}.  Inspired by the experimental results, some design principles have been proposed for a chemical compass, e.g. freeing one radical pair of its nuclear spin environment (called reference-and-probe model) \cite{Ritz09,Kim08}, applying gradient fields \cite{Cai11} or by exploiting quantum criticality \cite{Sun11}. It is however unknown what is the sensitivity limit that a chemical compass can achieve,  whether and how quantum coherence can be exploited to achieve maximal magnetic sensitivity.

In this paper, we first investigate what kind of hyperfine coupling in a chemical compass can lead to better magnetic field sensitivity. Motivated by our numerical optimization results, we focus on the simplest model configuration with one nuclear spin, which appears to offer the best sensitivity. This simple model is an approximation of multiple nuclei environments with one effective dominant coupling component. It does also allow us to obtain analytic expressions for the optimized magnetic sensitivity so that we can better understand the underlying physics. A strong anisotropic hyperfine coupling has usually been considered to give a high sensitivity. Interestingly, we find a very rich structure in the sensitivity as a function of the hyperfine coupling, which has not been expected before. We show that a chemical compass with a reasonable radical pair lifetime can in principle work for a magnetic field as weak as a few tens of nano Tesla ($nT$). Furthermore, we demonstrate that the sensitivity limit can be extended with simple quantum control, i.e. using an extra time-dependent magnetic field to control the radical pair dynamical processes. This result illustrates how quantum coherence can be exploited to improve the sensitivity of a chemical compass. Decoherence, coming from the interaction with the environment (out of the core chemical system of the radical pair and the adjacent nuclei), is unavoidable at room temperature and is generally regarded to exert a destructive influence on a chemical compass. Perhaps surprisingly, we demonstrate that the performance of a chemical compass can be very robust against dephasing and can even be improved by increasing the dephasing rate. This phenomenon clearly shows that an interplay between coherent spin dynamics of radical pairs and environmental decoherence can play an interesting role in a chemical compass.

{\it Optimization of a chemical compass.---} In a chemical compass, each radical has an unpaired electron coupled to an external magnetic field $\vec{B}$ and a few nuclei via the Hamiltonian \cite{Ste89} $H=-\gamma _{e} \vec{B}(\vec{S}_{A}+\vec{S}_{D})+\sum_{k=A,D}\sum_{j}\vec{S}_{k}\cdot \hat{T}_{k_{j}}\cdot \vec{I}_{k_{j}}  $,where $\gamma _{e}=-g_{e}\mu _{B}$ is the electron gyromagnetic ratio, $\hat{T}_{k_{j}}$ denote the hyperfine coupling tensors, and $\vec{S}_{k}$ and $\vec{I}_{k_{j}}$ are the electron and nuclear spin operators, respectively. For simplicity, we assume that dipole-dipole and exchange interactions between two radicals can be neglected. This approximation is valid when the radical-radical distance is relatively large or the dipole-dipole and exchange interactions partially cancel each other, as proposed in Ref. \cite{Efi08}. The hyperfine coupling is the most essential ingredient of a chemical compass, as it leads to the coherent interconversion between the electron spin singlet and triplet states, which then recombine into different chemical products at a certain rate.  It has not been fully understood what properties it should have to enable a chemical compass to have maximal magnetic sensitivity.

We consider a radical pair created in a spin-correlated electronic singlet state $|S\rangle =\frac{1}{\sqrt{2}}(\left\vert \uparrow\downarrow \right\rangle -\left\vert \downarrow \uparrow\right\rangle )$ with the same singlet and triplet recombination rates, i.e. $k_{S}=k_{T}=k$. Under ambient conditions, the nuclear spins initially are assumed to be at thermal equilibrium, i.e. described by a density matrix as $\bigotimes_{j}\mathbb{I}_{j}/d_{j}$, where $d_{j}$ is the dimension of the $j$th nuclear spin and $\mathbb{I}_{j}$ is the identity matrix. The singlet yield calculated from the conventional phenomenological approach \cite{Ste89} is quite similar to the one from the quantum-measurement-based master equation \cite{Kom09JonHor10}, but requires much less computational efforts. Explicitly, it is calculated as $\Phi_{S}=\int_{0}^{\infty}r(t)f_{S}(t)dt$, where $r(t)=k \exp{(-k t)}$ is the radical recombination probability distribution, and $f_{S}(t)$ is the singlet ratio of the electron spin state $\rho_{s}(t)$ at time $t$, i.e. $f_{S}(t)=\langle S | \rho_{s}(t) | S \rangle$.

As a known design principle of a chemical compass, partially inspired by the behavior experiments with Europe robins \cite{Ritz04}, it was proposed that in an optimally designed chemical compass, only one of the radicals has strong and anisotropic hyperfine interactions, while the other is free from the coupling with nuclear spins \cite{Ritz09,Kim08}. Thus, we take this kind of reference-and-probe model as the starting point for our optimization procedure. The target function is the magnetic sensitivity of the singlet yield quantified by
\begin{equation}
D_S=\Phi_S^{(\mbox{max})}-\Phi_S^{(\mbox{min})} \; ,
\end{equation}
where $\Phi_S^{(\mbox{max})}$ and $\Phi_S^{(\mbox{min})}$ are the maximum and minimum singlet yield for different directions of the magnetic field. In this scenario our numerical optimization of the hyperfine couplings suggests that a chemical compass having more nuclear spins will not perform better than having only one nuclear spin \cite{TNS}. We thus consider the simple case of one nuclear spin with the anisotropic hyperfine coupling tensor as $\hat{T}=\mbox{diag}\{ 0,0,a\}$ that can give the best sensitivity. Such a simple model allows us to obtain analytic results which are quite useful for understanding the essential mechanism responsible for a chemical compass. The real molecule for avian magnetoreception would be more complicated than this model, while the physics revealed by the simple model may still provide insights into avian magnetoreception. In fact, it can be an approximation of more general models with one dominant hyperfine component. It may also be possible that the motion/disorder of protein may average the complicated interactions, and may eventually result in a simple effective Hamiltonian \cite{Schu}.
\begin{figure}[t]
\begin{center}
\hspace{-0.5cm}
\includegraphics[width=8.5cm]{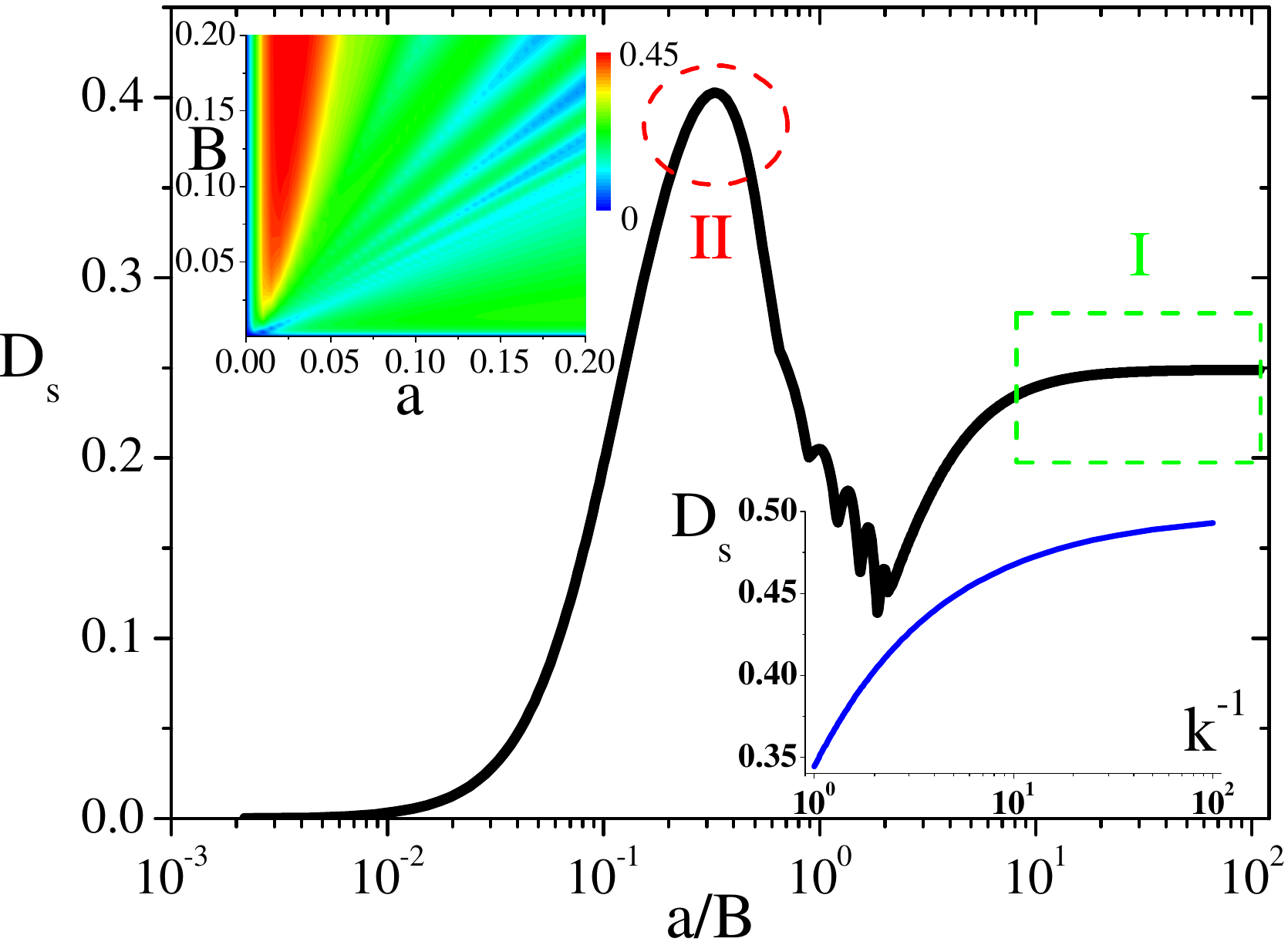}
\end{center}
\caption{(Color online) The magnetic sensitivity of the singlet yield, $D_S$, as a function of the ratio between hyperfine coupling $a$ and magnetic field strength $B$. The radical pair recombination rate is $k=0.5\mu s^{-1}$, and the magnetic field is set as geomagnetic field $B=46\mu T$. A chemical compass can have a good sensitivity if the hyperfine coupling lies in two regimes, see the rectangle I and the ellipse II. Inset (top left): contour plot for $D_S$ versus $a$ and $B$, in the case of $k=0.5\mu s^{-1}$ as above. Inset (bottom right): $D_S$, achieved by the optimal $a$, as a function of the radical pair lifetime $\tau=k^{-1}$ (in $\mu s$). }
\end{figure}

The direction of the weak magnetic field $\vec{B}$ is characterized by $(\theta,\phi)$. Without loss of generality, we can assume that $\phi=0$. The Hamiltonian then simplifies to $H=\vec{B}\cdot \vec{S}_1+\vec{B}\cdot \vec{S}_2+a S_2^z I_z$. The nuclear spin polarization remains unchanged during the spin dynamics. This is also true if the transverse component of the hyperfine coupling is small and thereby the spin flip is prohibited due to the energy mismatch. The nuclear spin can then be treated as inducing an effective magnetic field (dependent on its state) for the electron spin. If the nuclear spin is in the up (down) state $\ket{\uparrow}_z$ ($\ket{\downarrow}_z$), the effective magnetic filed is $\frac{1}{2} a \hat{z}$ ($-\frac{1}{2} a \hat{z}$). Thus, we can written the effective Hamiltonian as $H_e=\vec{B}\cdot \vec{S}_1+\vec{B}\cdot \vec{S}_2+b S_2^z$ where $b=\pm\frac{1}{2} a $. After some calculations, we obtain the singlet yield as a function of $\theta$ and $a$ as
$\Phi_S(\theta,b)=\frac{1}{4}\bla{1+c^2}+\frac{1}{4}\bla{1-c^2}\blb{g(B_1)+g(B)}
+\frac{1}{8}\bla{1-c}^2 g(B_1+B) +\frac{1}{8}\bla{1+c}^2 g(B_1-B)$,
where $c=\cos(\theta-\theta')$, $g(x)=k^2/\bla{k^2+x^2}$, $B_1^2=(B\cos{\theta}+b)^2+B^2\sin^2{\theta}$, $\sin{\theta'}=B\sin{\theta}/B_1$, $\cos{\theta'}=(B\cos{\theta}+b)/B_1$. Therefore, the total singlet yield for the magnetic with an angle $\theta$ is
\begin{equation}
\Phi_S(\theta)=\frac{1}{2}\blb{\Phi_S(\theta,\frac{a}{2})+\Phi_S(\theta,-\frac{a}{2})} \; .
\end{equation}
This analytic result for the singlet yield enables us to have a better understanding when a chemical compass can (not) have a good magnetic sensitivity. For instance, if the recombination rate $k$ is too large, the radical pair will enter a quantum-Zeno regime and hence remain in the singlet state due to suppression of singlet-triplet transition \cite{Kom09JonHor10}. As a consequence, the singlet yield $\Phi_S(\theta) \simeq 1$ is almost independent on the value of angle $\theta$. A good sensitivity requires a moderate recombination rate to give enough time for magnetic field effects to occur without being severely degraded by spin decoherence.

As verified by our numerical optimization results, we find that a chemical compass can work with a high and robust sensitivity if the hyperfine coupling lies in one of two regimes, see Fig.1. In the first regime (I), the hyperfine coupling is relatively large, i.e. $a\gg B$. In this case, we have the singlet yield as
\begin{equation}
\Phi_S(\theta)\simeq \frac{1}{4}[1+\cos^2(\theta-\theta')]\simeq \frac{1}{4}\bla{1+\cos^2{\theta}} \; .
\end{equation}
Therefore, the maximum and minimum singlet yields are $\Phi_S^{(\mbox{max})}=\Phi_S(0,a)=\frac{1}{2}$, which corresponds to the coherent mixing between the singlet state $\ket{S}$ and the triplet state $\ket{T_0}=\frac{1}{\sqrt{2}}(\left\vert \uparrow\downarrow \right\rangle +\left\vert \downarrow \uparrow
\right\rangle )$, and $\Phi_S^{(\mbox{min})}=\Phi_S(\frac{\pi}{2},a)=\frac{1}{4}$ corresponding to the mixing between the singlet state and all three triplet states. Thus, the magnetic sensitivity is $D_S=\frac{1}{4}$. This is exactly the sensitivity that can be achieved by a chemical compass which has one of the nuclear spin with a large dominant anisotropic coupling along a certain direction. For natural molecules which do not satisfy such conditions of hyperfine couplings, one can use magnetic nano-particles to design a metal-chemical hybrid compass and achieve the above sensitivity \cite{Cai11}. For very weak magnetic fields, the magnetic sensitivity is $D_S(B)=B^2/\blb{4\bla{k^2+B^2}}$. This explicitly shows the dependence of the magnetic sensitivity on the the recombination rate $k$ (i.e. the radical pair lifetime). Let us point out that the other benefit of having a large local effective field is to make the functioning of a chemical compass more robust against decoherence. In the second regime (see II in Fig.1) , the hyperfine coupling is smaller than the magnetic field, and the interplay between them leads to an even better sensitivity ($D_S=0.4$ for $k=0.5\mu s^{-1}$ and $a=\frac{1}{3}B$). The maximum singlet yield is obtained at  $\theta=\frac{\pi}{2}$, where the hyperfine coupling is not large enough to cause the transitions from the singlet state to the triplet states (as $a<B$); while the minimum singlet yield happen at $\theta=0$ that corresponds to the coherent mixing between the singlet state $\ket{S}$ and the triplet state $\ket{T_0}$. Let us stress that an optimally designed chemical compass can achieve a sensitivity more than $10 \%$ even for a magnetic field that is one order magnitude weaker than the geomagnetic field (with the radical pair lifetime $2\mu s$) - see inset in Fig.1. In fact, we find that a chemical compass may work at a magnetic field down to a few tens of nanotelsa if the radical pair life time is reasonably longer (e.g. $\sim 10 \mu s$). This may be of practical interest if one wants to find applications for an artificial chemical compass. Finally, the robustness of such optimal configurations is shown in the contour-plot inset of Fig.1, i.e. the best performance is obtained in a relatively large range of the values of $a$ and $B$.

{\it Extend magnetic sensitivity limit with simple quantum control.---}  As an application of quantum control in the context of radical pair reactions, it was shown that quantum dynamical coupling, which is a very specific kind of quantum control, can be used to disturb the function of a chemical compass. The essence is to decouple the hyperfine coupling and thus malfunction the compass \cite{Cai10prl}. The role of this special quantum control is {\it destructive}. In contrast, here we show that by simple quantum control we can enhance the sensitivity of a chemical compass and go beyond the sensitivity limit with the optimal hyperfine coupling found above. We introduce an external control magnetic field (in the direction close to the $\hat{x}$ axis) that is a simple function of time as $C(t)=\sum_{k} A_k \sin{(\omega_k t)}+B_k \cos{(\omega_k t)}$ under some appropriate amplitude and spectral constraints, which depend on the experimental conditions. Similar techniques have been recently proposed theoretically to prepare generic coherent or incoherent initial states of light harvesting systems \cite{Caruso11}.
\begin{figure}[t]
\begin{center}
(a) \hspace{4.2cm} (b)
\end{center}
\vspace{-1.1cm}
\begin{center}
\begin{minipage}{9cm}
\hspace{-0.8cm}
\includegraphics[width=4.6cm]{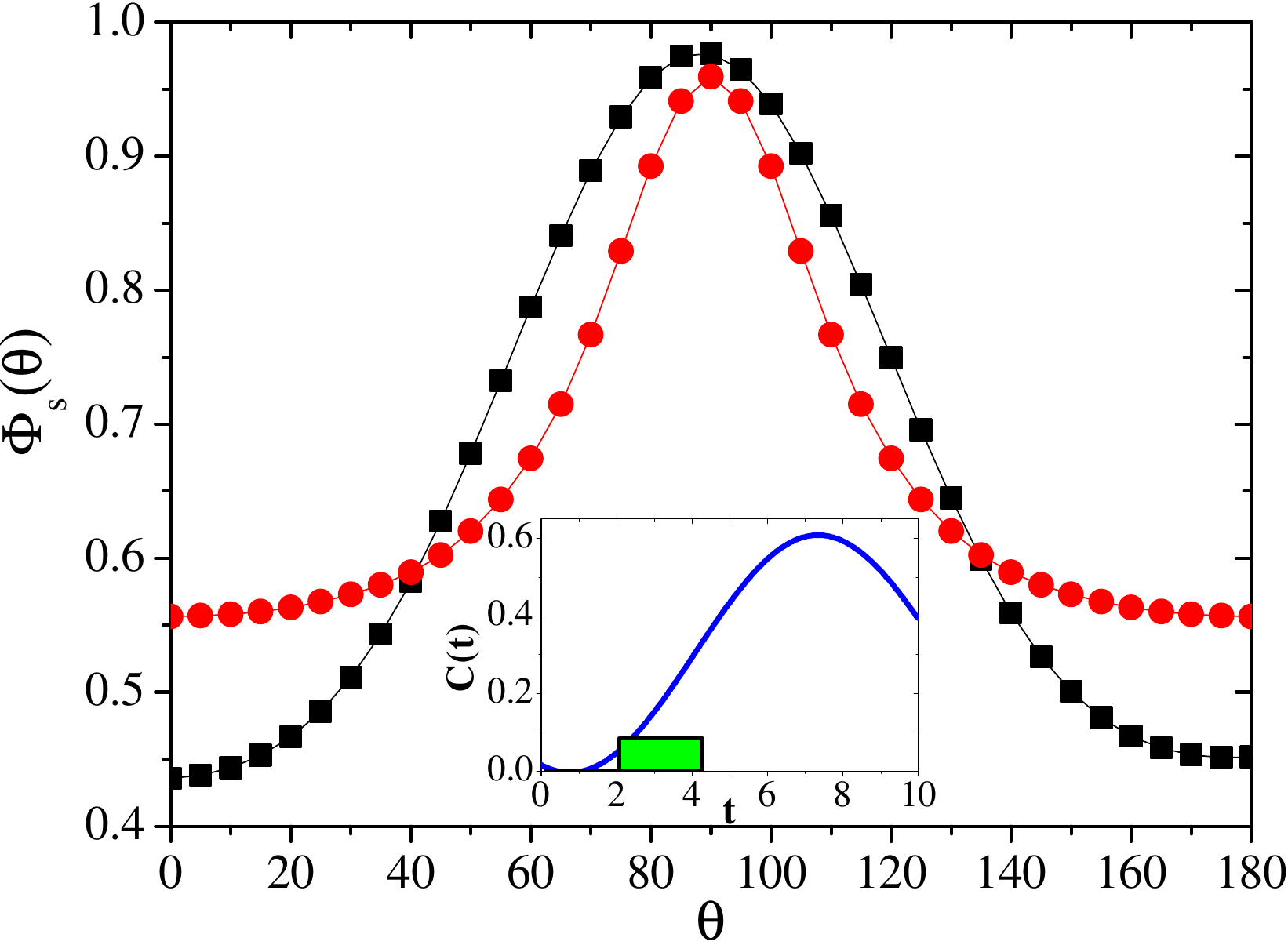}
\hspace{-0.2cm}
\includegraphics[width=4.6cm]{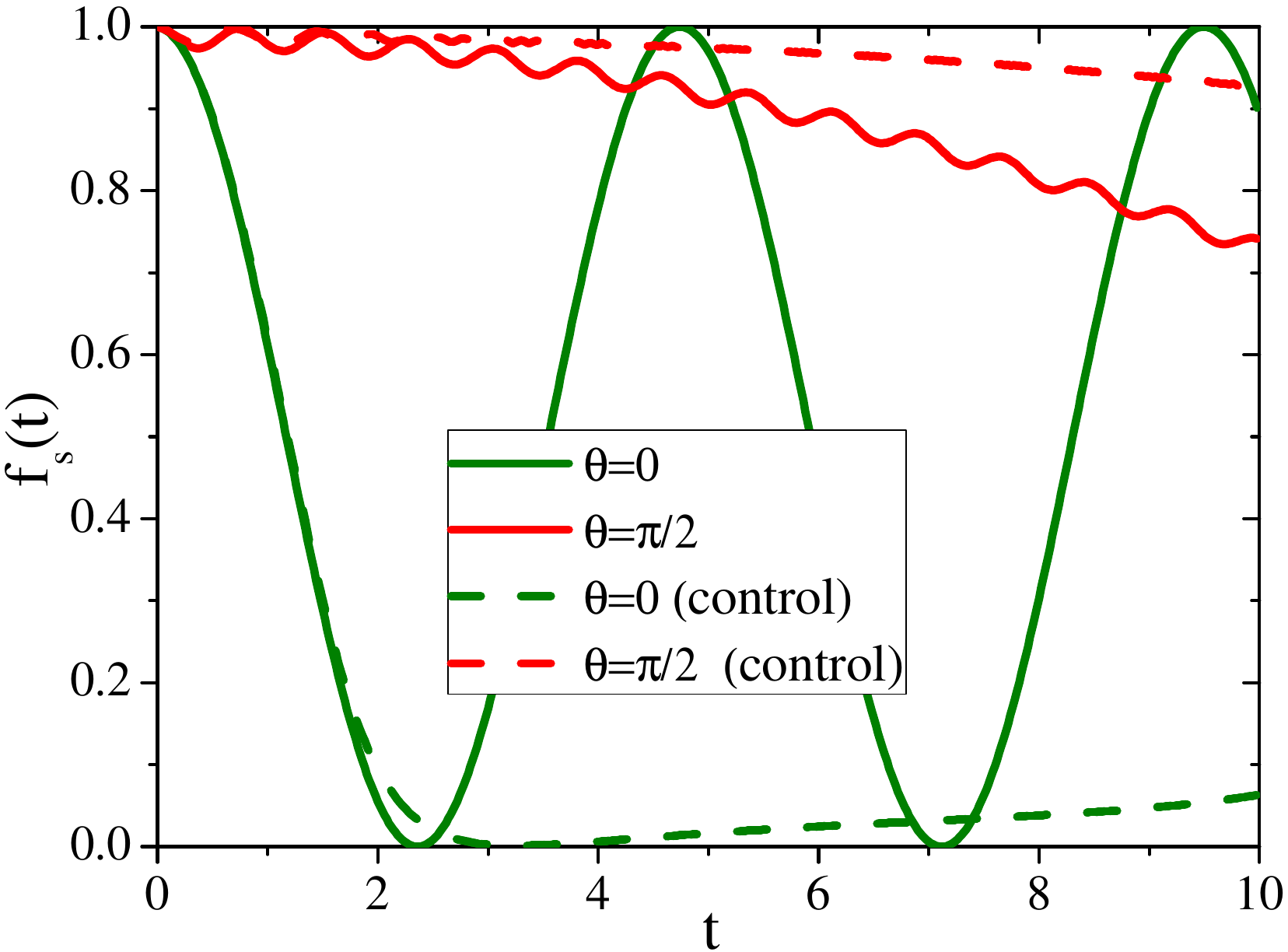}
\end{minipage}
\end{center}
\caption{(Color online) (a) The singlet yield $\Phi_s(\theta)$ as a function of the angle $\theta$ without (black square) and with (red circle) quantum control. Inset: simple control magnetic field $C(t)$ (in $mT$) versus time ($\mu s$). A much weaker and shorter step function (green area) for $C(t)$ allows us to achieve a very similar sensitivity ($\sim 53 \%$) as the optimal one ($\sim 54 \%$), further showing the robustness of such external control. (b) The singlet ratio $f_s(t)$ for $\theta=0$ (green) and $\theta=\frac{\pi}{2}$ (red), versus time (in $\mu s$), with (dashed) and without (solid) quantum control.} \label{OPC1}
\end{figure}
Our numerical optimizations, based on various algorithms as Nelder-Mead simplex direct search and subplex methods for unconstrained problems, shows that quantum control can improve the magnetic sensitivity significantly - see Fig.2(a). For a chemical compass with the optimal hyperfine coupling, the sensitivity $D_S$ (for the radical pair lifetime $k^{-1}=2\mu s$) is improved from $0.4$ to $0.54$, which is even better than the sensitivity limit $0.5$ (that can be obtained only in the long radical pair lifetime limit without quantum control). For simplicity, we use two frequencies of oscillating fields with moderate amplitudes in order to generate realistic control fields as in the inset of Fig.2(a). The result is quite surprising, as for other magnetic sensors the sensitivity tends to be degraded if one applies an additional magnetic field which may overwhelm the directional information of the magnetic field that we want to detect. However, in the current case, the appropriate applied field (on both spins, and this makes it essentially different from the method with gradient fields) can instead amplify the directional effect of the weak magnetic field on the singlet yield. The phenomenon can be understood by looking at the evolution of the singlet yield ratio over time  for $\theta=0$, see Fig.2(b). In the first 2 $\mu s$, the control field is very small, the singlet state $\ket{S}$ coherently transfers to the triplet state $\ket{T_0}$, and then the increasing control field suppresses the back-conversion from $\ket{T_0}$ to $\ket{S}$ (by inducing rapid transitions among the three triplet states) which reduces the minimum singlet yield. In the mean time, the control field will keep the singlet state at $\theta=\frac{\pi}{2}$, leading to the maximum singlet yield. The above essential physics suggests that we can use other forms of control field (dependent on specific experimental conditions, which may not take the form of oscillating fields) to achieve a similar and robust sensitivity improvement. Indeed, motivated by this intuition, we have also applied a much shorter and weaker step-like magnetic field achieving essentially the same sensitivity - see inset Fig.2(a). We remark that optimal control techniques can be exploited in a more general chemical compass with several nuclear spins to achieve a very high magnetic sensitivity enhancement, e.g. $60\%$ for the FADH$^\cdot$-O$_2^{\cdot -}$ molecule \cite{Hore2003,Sol09}.

The optimization approach we use here can be also translated to design experiments with open-loop feedback control in order to achieve different targets, e.g. controlling the chemical yields and the radical pair lifetime. Once we gain more knowledge about different specific radical pair reactions from this type of spin chemistry experiments, it will be interesting to apply the obtained optimal control results to experiments on the actual biological organisms, also verifying whether they really contain such investigated molecules.

{\it Decoherence-enhanced magnetic sensitivity of a chemical compass.---} The most prominent feature of avian magnetoreception is that it can work at room temperature when various kinds of noise may exist. Under such conditions, quantum systems (here radical pair spins) suffer from decoherence and quantum features will be weaken. One will naturally expect that quantum decoherence will be unfavorable for the functioning of a chemical compass. Indeed, it was shown that if the decoherence rate is larger than the recombination rate, the magnetic sensitivity of a model compass will be dramatically degraded by general noises, while for a particular dephasing model (which is different from the model we study here), the singlet yield itself is entirely unaffected \cite{Ved09}. Here, we show that a chemical compass with an optimal hyperfine coupling will not only have a good sensitivity, but is also quite robust against pure dephasing, and surprisingly, stronger dephasing may also enhance its magnetic sensitivity. For simplicity, we model noise as pure local dephasing described by the following Lindblad operators \cite{Caruso09} $\mathcal{L}_P (\rho) = \frac{1}{4}\sum\limits_{k=1,2} \bla{2 L_k \rho L_k^{\dagger}-L_k^{\dagger} L_k \rho-\rho L_k^{\dagger} L_k}$ with $L_1=\bla{\frac{\gamma}{1+d^2}}^{1/2} \blb{\sigma_z^{(1)}+d\sigma_z^{(2)}}$ and $L_2=\bla{\frac{\gamma}{1+d^2}}^{1/2} \blb{d\sigma_z^{(1)}+\sigma_z^{(2)}}$, where $\gamma$ is the dephasing rate and $\sigma_z$ is Pauli operator. This quantum master equation approach is equivalent and can be directly mapped to a Liouville equation, see e.g. \cite{Hore11}. The parameter $d$ characterizes how correlated the dephasing is, i.e. $d=0$ for uncorrelated dephasing and $d=1$ for perfectly correlated one. We note that, since the distance between two radicals is of the order of several nanometers, the local energy fluctuations and thus dephasing may well be correlated. In Fig.3, we plot the magnetic sensitivity as a function of the dephasing rate $\gamma$. It can be seen from Fig.3(a) that the performance of such a chemical compass is quite robust and even enhanced by the presence of correlated dephasing.  For the uncorrelated dephasing, the sensitivity will first be degraded but will be enhanced again as the dephasing rate further increases. This result comes from the interplay between the coherent spin Hamiltonian and the noise, which is evidenced in Fig.3(b). A similar phenomenon happens for the anti-correlated dephasing. Let us remark that, if the dephasing rate is extremely large and overwhelms all the coherent times scales, the spin dynamics is simply incoherent mixing of $\ket{S}$ and $\ket{T_0}$, and the sensitivity will eventually be degraded as expected.

\begin{figure}[t]
\begin{center}
(a) \hspace{4.2cm} (b)
\end{center}
\vspace{-1.0cm}
\begin{center}
\begin{minipage}{9cm}
\hspace{-0.8cm}
\includegraphics[width=4.6cm]{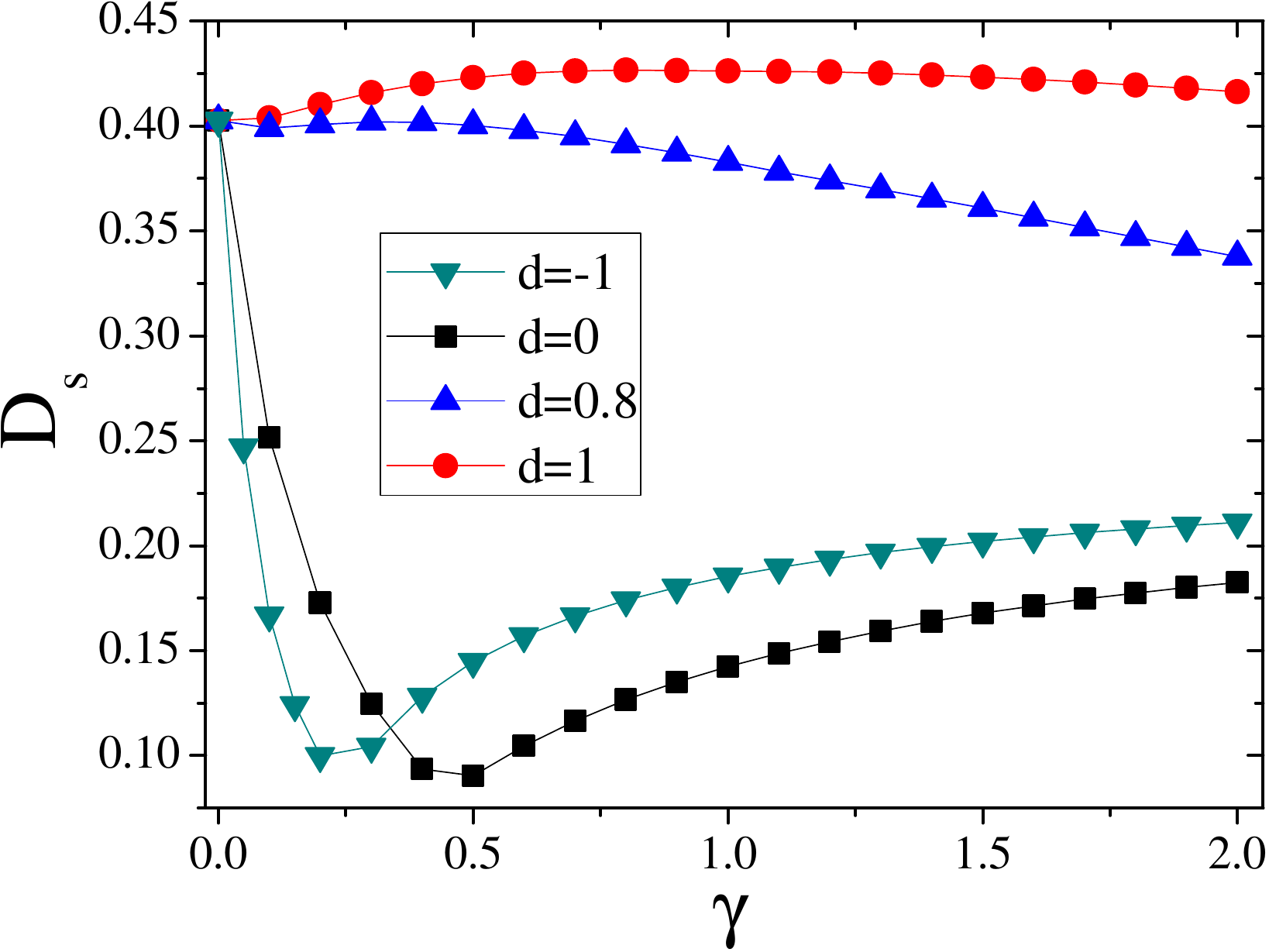}
\hspace{-0.1cm}
\includegraphics[width=4.6cm]{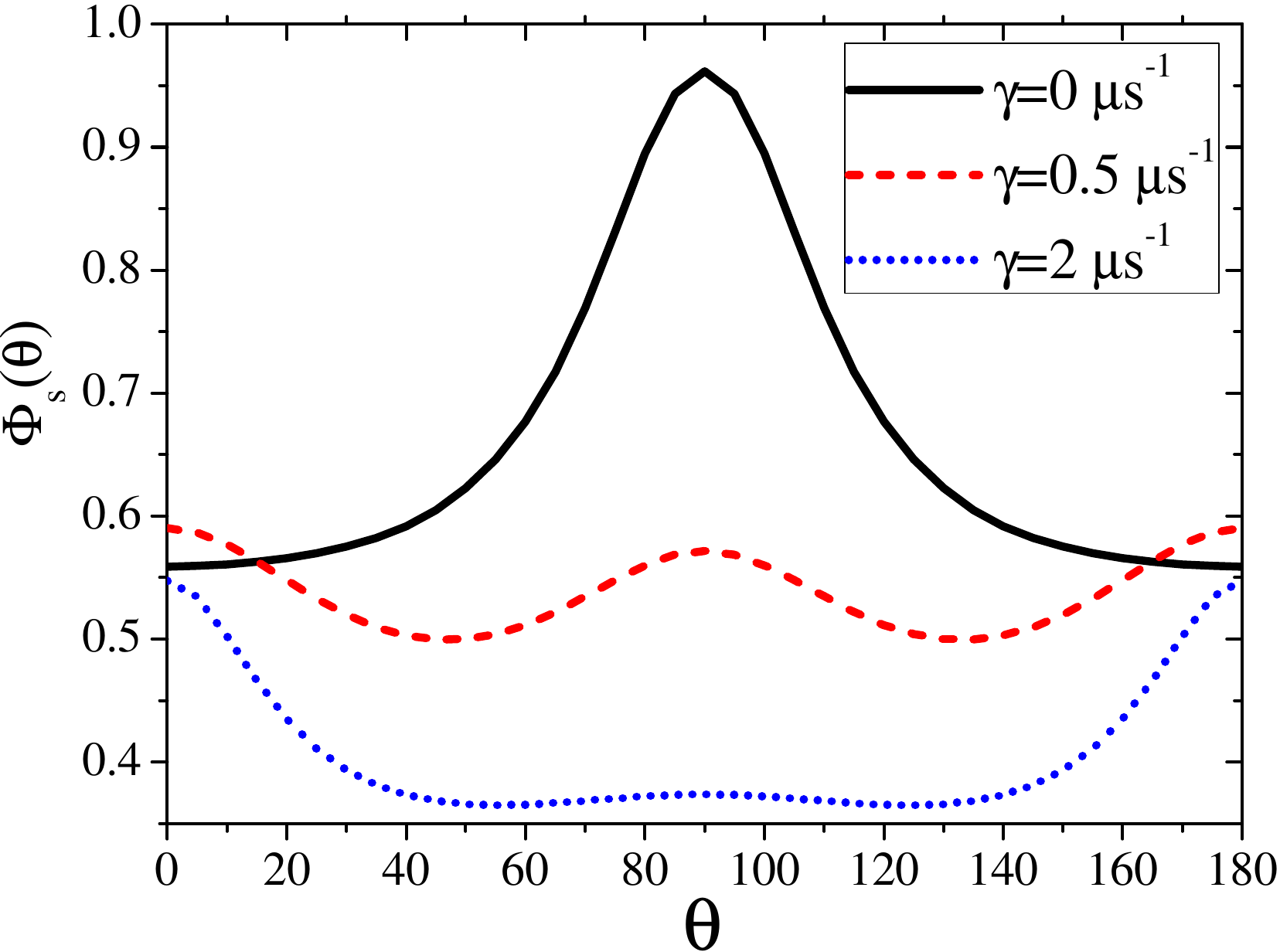}
\end{minipage}
\end{center}
\caption{(Color online) (a) The magnetic sensitivity of the singlet, $D_S$,  as a function of the dephasing rate $\gamma$ ($\mu s^{-1}$). The radical pair recombination rate is again $k=0.5 \mu s^{-1}$. It shows the effect of uncorrelated ($d=0$), partially correlated ($d=0.8$), perfectly correlated ($d=1$), and anti-correlated ($d=-1$) dephasing on $D_S$ for an optimally designed chemical compass. (b) Singlet yield $\Phi_S(\theta)$ for uncorrelated dephasing with rates $\gamma=0$, $0.5$, $2\mu s^{-1}$, respectively.}
\end{figure}

{\it Summary.---} We investigate quantum limits for the magnetic sensitivity of a chemical compass by optimizing its hyperfine coupling. We find that a model chemical compass with only one nuclear spin equipped by an optimal hyperfine coupling is already likely to give us a very high magnetic sensitivity. For such an optimally designed chemical compass, we demonstrate that its sensitivity can be further improved with simple quantum control. Moreover, its performance can be very robust and even better if the noise is correlated. In the case of uncorrelated dephasing, the magnetic sensitivity does not monotonically decrease with the dephasing rate. These phenomena stem from the intricate interplay between quantum coherence and unavoidable noise. Therefore, our results may help us to understand what quantum coherent features of radical pair mechanism lead to chemical compass sensitivity, and provide useful design principles for a biologically inspired artificial chemical compass.

{\it Acknowledgements} The work was supported by the Alexander von Humboldt Foundation and by the EU STREP project CORNER. F.C. and J.-M.C were supported also by a Marie-Curie Intra-European Fellowship within the 7th European Community Framework Programme.


\begin{thebibliography}{99}
\bibitem{TNS} Here we focus on the model that one radical is free of hyperfine coupling. In fact, if we allow both radicals to couple with nuclear spins and optimize the hyperfine couplings, it is possible to have only a slightly better sensitivity. We will address this scenario in detail elsewhere.
\end{thebibliography}
\end{document}